\documentclass[letterpaper,titlepage,11pt]{article}
\usepackage[usenames,dvipsnames]{xcolor}

\usepackage{amssymb,amsmath,amsfonts}

\usepackage[colorlinks,hyperindex, % pagebackref,
            bookmarks=true,bookmarksopen=true]{hyperref}
    \hypersetup
    {
        colorlinks=true,
        linkcolor=MidnightBlue,
        urlcolor=MidnightBlue,
        filecolor=black,
        citecolor=RedOrange,
        pdfstartview=FitV,
        pdftitle={},
        pdfauthor={Ilmar Gahramanov, Grigory Vartanov},
        pdfsubject={},
        pdfkeywords={},
        pdfpagemode=None,
        bookmarksopen=true
    }

\usepackage{mathtext}
\usepackage{stackrel}

\begin{document} 

\begin{titlepage}

\rightline{\vbox{\small\hbox{\tt DESY 13-041} }}
\rightline{\vbox{\small\hbox{\tt HU-EP-13/11} }}
 \vskip 2.7 cm

\centerline{\Large \bf Extended global symmetries for 4d ${\cal N}=1$ SQCD theories}
\vskip 1 cm

\centerline{\large {\bf Ilmar Gahramanov\footnote{ilmar.gahramanov@physik.hu-berlin.de}$\,^{a,b,c}$} and
{\bf Grigory Vartanov\footnote{grigory.vartanov@desy.de}$\,^{a}$} }

\begin{center}
\textit{$^{a}$  DESY Hamburg, Theory Group,\\ Notkestrasse 85, D-22607 Hamburg, Germany} \\
\texttt{} \\
\vspace{.2mm}
\textit{$^{b}$ Institut f\"{u}r Physik, Humboldt-Universit\"{a}t zu Berlin,\\ 
Newtonstrasse 15, 12489 Berlin, Germany} \\
\texttt{} \\
\vspace{.2mm}
\textit{$^{c}$ Institute of Radiation Problems ANAS,\\ B.Vahabzade 9, AZ1143 Baku, Azerbaijan} \\
\texttt{} 
\vspace{0mm}
\end{center}

\vskip 1.5cm \centerline{\bf Abstract} \vskip 0.2cm \noindent In \href{http://arxiv.org/abs/arXiv:0811.1909}{arXiv:0811.1909} Spiridonov and Vartanov, using the superconformal index technique, found that 4--dimensional ${\cal N}=1$ SQCD theory with $SU(2)$ gauge group and four flavors has $72$ dual representations. Recently in \href{http://arxiv.org/abs/arXiv:1209.1404}{arXiv:1209.1404} the authors showed that these dual theories, when coupled to $5d$ hypermultiplets with specific boundary conditions have an extended $E_7$ global symmetry. In this work we find that for a reduced theory with $3$ flavors the explicit $SU(6)$ global symmetry is enhanced to an $E_6$ symmetry in the presence of $5d$ hypermultiplets. We also show connections between indices of different theories in $3$ and $4$ dimensions.

\end{titlepage}

\small
\tableofcontents
\normalsize
%\newpage
%\pagestyle{plain}
\setcounter{page}{1}

\section{Introduction and conclusions}
\label{sec:intro}
In recent years considerable progress has been made in the study of rigid supersymmetric field theories in nontrivial spacetimes. In particular the superconformal index has been a primary goal of the recent interest in these theories. The index is a powerful tool to test Seiberg--like dualities in ${\cal N}=1$ \cite{Romelsberger:2007ec, Dolan:2008qi, Spiridonov:2008zr, Spiridonov:2009za, Spiridonov:2011hf}, S--dualities in ${\cal N}=2$ \cite{Gadde:2009kb, Gadde:2011uv} and ${\cal N}=4$ \cite{Gadde:2009kb, Spiridonov:2010qv} supersymmetric theories and has an elegant mathematical structure described by the theory of elliptic hypergeometric integrals \cite{S1,S2}. 

The superconformal index was introduced \cite{Romelsberger:2007ec, Romelsberger:2005eg, Kinney:2005ej} as a nontrivial generalization of the Witten index \cite{Witten:1982df}, which counts BPS states in superconformal field theories in curved spacetime \cite{Festuccia:2011ws}. We give a short outline of a superconformal index and refer the reader to \cite{Spiridonov:2008zr, Spiridonov:2009za, Sudano:2011aa} for more details. 

Let us consider the ${\cal N}=1$ superconformal theory in four dimensions. The symmetry group of this theory is $SU(2,2|1)$, which has the following generators: Lorentz rotations $J_i$, $\bar{J}_i$, supertranslations $P_\mu$, $Q_{\alpha}$, $\bar{Q}_{\dot{\alpha}}$ with $\{Q_\alpha, \bar{Q}_{\dot{\alpha}}\}=2 P_{\alpha \dot{\alpha}}$, special superconformal transformations $K_\mu$, $S_\alpha$, $S_{\dot{\alpha}}$ with $\{\bar{S}^{\dot{\alpha}},S^\alpha \}=2 K^{\dot{\alpha} \alpha}$, dilatations $H$   and $U(1)_R$--rotations $R$.
To construct the superconformal index let us consider, for example, the supercharges $\bar{Q}_1$ and $\bar{S}^1$, which satisfy the following relation
\begin{equation}
\{\bar{Q}_1, \bar{S}^1\}=-2(H-2\bar{J}_3-\frac32 R) \; .
\end{equation}
Then one defines the superconformal index in the following way
\begin{equation}\label{Ind}
{\rm ind}(t,x,g,f) \ = \ Tr \left( (-1)^{\rm F}x^{2J_3}t^{\mathcal{R}}
e^{\sum_{a=1}^{rank\, G} g_aG^a} e^{\sum_{j=1}^{rank\, F} f_jF^j}
\right) \;.
\end{equation}
Here $(-1)^F$ is the fermion number operator, $t^{\mathcal{R}}$ and $x^{2J_3}$ are additional regulators with $|t|<1$ and $|x|<1$, $g_a$ and $f_j$ are the chemical potentials for groups $G$ and $F$ respectively, where $G$ is a non-abelian gauge group with maximal torus generators $G_a,\, a=1,\ldots,$ rank $G$, and $F$ is a flavor group  with maximal torus generators $F_j,\, j=1,\ldots,$ rank $F$. 

According to the Romelsberger prescription \cite{Romelsberger:2007ec} for ${\cal N}=1$ superconformal theories one can write the full index via a ``plethystic'' exponential \cite{Feng:2007ur} and integrate over the gauge group\footnote{Because we are interested in gauge invariant physical observables.}
\begin{equation} \label{plethystic}
I(p,q,\underline{y})  =  \int_{G_c} d \mu(g)\, \exp \bigg ( \sum_{n=1}^{\infty}
\frac 1n \text{ind}\big(p^n ,q^n, \underline{z}^n , \underline{y}^ n\big ) \bigg ),
\end{equation}
where $d \mu(g)$ is the $G_c$--invariant measure and the single particle states index is
\begin{align} 
\text{ind}(p,q,\underline{z},\underline{y}) &= \frac{2pq - p - q}{(1-p)(1-q)} \chi_{adj}(\underline{z})  \nonumber \\ \label{singleindex}
& \quad + \sum_j \frac{(p q)^{R_j/2}\chi_{R_F,j}(\underline{y})\chi_{R_G,j}(\underline{z}) - (pq)^{1-R_j/2}\chi_{{\bar R}_F,j}(\underline{y})\chi_{{\bar R}_G,j}(\underline{z})}{(1-p)(1-q)}.
\end{align}
Here we introduced the new parameters $p=t x$ and $q=t x^{-1}$. The first term in (\ref{singleindex}) represents the contribution of the gauge superfields lying in the adjoint representation of the gauge group $G_c$. The sum over $j$ corresponds to the contribution of chiral matter superfields $\varphi_j$ transforming in the gauge group representations $R_{G,j}$ and flavor group representations $R_{F,j}$ where $R_j$ are the field $R$-charges. The functions $\chi_{adj}(\underline{z})$, $\chi_{R_F,j}(\underline{y})$ and $\chi_{R_G,j}(\underline{z})$ are the  characters of the corresponding representations, where $\underline{z}$ and $\underline{y}$ are the set of complex eigenvalues of matrices realizing $G$ and $F$, respectively.

Dolan and Osborn realized \cite{Dolan:2008qi} that the exponential sum in (\ref{plethystic}) can be evaluated using elliptic Gamma function
\begin{equation}
\Gamma(z;p,q)= \prod_{i,j=0}^\infty
\frac{1-z^{-1}p^{i+1}q^{j+1}}{1-z p^i q^j}, \quad |p|, |q|<1 \; ,
\end{equation}
and as a result the superconformal index can be expressed in terms of Spiridonov's elliptic hypergeometric integrals. For a detailed discussion, see \cite{Spiridonov:2009za} and also \cite{Spiridonov} for mathematical aspects of these integrals. Note that in the rest of the paper we will use the following standard shorthands
\begin{align}
\Gamma(z,w;p,q)& \equiv \Gamma(z;p,q)\Gamma(w;p,q),\\
\Gamma(z^{\pm k};p,q)& \equiv \Gamma (z^k;p,q) \Gamma(z^{-k};p,q).
\end{align}

In \cite{Spiridonov:2008zr} authors established multiple dualities based on the so--called $V$--function
\begin{equation}
    I(t_1, \ldots , t_8;p,q) = \frac{(p;p)_{\infty} (q;q)_{\infty}}{2} \int_{{\mathbb T}}
\frac{\prod_{j=1}^8 \Gamma(t_j z^{\pm 1};p,q)}{\Gamma(z^{\pm 2};p,q)}
\frac{dz}{2 \pi i z}, 
\end{equation}
where $t_j,\; j=1,\ldots,8$ are complex parameters with the balancing condition $\prod_{j=1}^8t_j=(pq)^2$ and the q-Pochhammer symbol $(z;q)_\infty=\prod_{i=0}^\infty (1-z q^i)$. They speculated on existence of $E_7$ global symmetry of the $V$--function from the fact that it has $W(E_7)$ Weyl symmetry group for integral transformation. In fact this symmetry was realized explicitly, based on $4d/5d$ system, by Dimofte and Gaiotto in \cite{Dimofte:2012pd}. 

In \cite{Spiridonov:2008zr} the authors reduced $4d$ ${\cal N}=1$ SYM with $SU(2)$ gauge group with $8$ quarks to $6$ quarks and found that the index of the reduced theory has $W(E_6)$ symmetry. After this reduction in the dual theories they realized additional $SU(2)$ global symmetries, the appearance of which was unclear to the authors. In this work we give the explanation of this extended symmetry by coupling of original $N_f=3$ theory to free $5d$ hypermultiplets\footnote{Note that we use the subscript $F$ for the flavor and the subscript $f$ for the number of quarks.}. This coupling bring us to $E_6$ global  symmetry. At the same time this $E_6$ symmetry can be obtained by restricting two parameters in combined $4d/5d$ index considered by Dimofte and Gaiotto \cite{Dimofte:2012pd}. 

We have $E_6$ global symmetry group and in different phases it produces us additional $SU(2)$ or $U(1)$ groups in dualities found in \cite{Spiridonov:2008zr}.

Our aim is to show connections between indices of different theories. The following ``commutative diagram'' demonstrates the plan of the paper pictorially. In section 2 we describe the reduction procedure from $4d$ $N_F=4$ theory to $3d$ $N_f=4$, in the diagram it follows to the anticlockwise direction. In section 3 we do further reduction from $3d$ $N_f=6$ theory, which is right down arrow in the diagram.

{\small
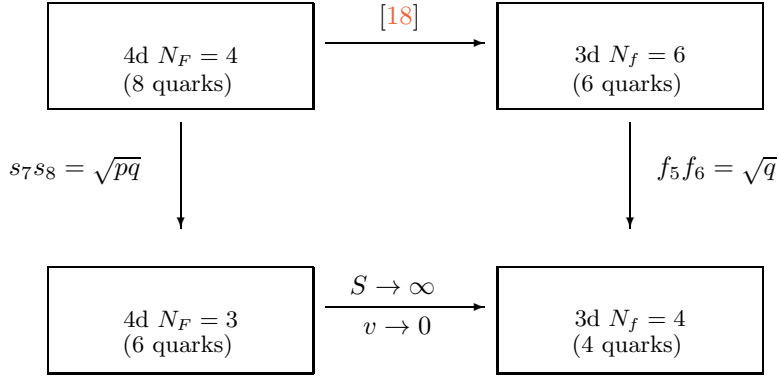
\begin{picture}(300,170)(-75,5)
\put(0,0){\framebox(100,40){{\footnotesize 4d $N_F=3$}}}
\put(28,8){{\footnotesize (6 quarks)}}
\put(0,100){\framebox(100,40){{\footnotesize 4d $N_F=4$}}}
\put(28,107){{\footnotesize (8 quarks)}}
\put(170,0){\framebox(100,40){{\footnotesize 3d $N_f=4$}}}
\put(198,8){{\footnotesize (4 quarks)}}
\put(170,100){\framebox(100,40){{\footnotesize 3d $N_f=6$}}}
\put(198,107){{\footnotesize (6 quarks)}}
\put(105,25){{\vector(1,0){60}}}
\put(114,30){$S\rightarrow \infty$}
\put(125,132){\cite{Dimofte:2012pd}}
\put(119,15){$v\rightarrow 0$}
\put(105,125){\vector(1,0){60}}
\put(50,95){{\vector(0,-1){40}}}
\put(-15,75){$s_7 s_8=\sqrt{p q}$}
\put(220,95){{\vector(0,-1){40}}}
\put(230,75){$f_5 f_6=\sqrt{q}$}
\end{picture}}

\vspace{0.6cm}

\noindent In this diagram $s_i$ and $f_i$ are the chemical potentials. The limit $v \rightarrow 0$ corresponds to dimensional reduction on the $S^1$ and $S\rightarrow \infty$ corresponds to the sending mass of quark supermultiplet to infinity. 

\section{Reduction of $4d$ SCI to $3d$ partition function}

In this section we will discuss $4d$ $SU(2)$ ${\cal N}=1$ SQCD theories with $N_F=4$. Let us consider first the electric theory with the flavor symmetry group $SU(8)$. The superconformal index for this theory is\footnote{This is the so--called $V$--function.}
\begin{align} \label{inindex}
I_{4d,\; N_F=4} =\frac{(p;p)_\infty (q;q)_\infty}{2}
 \oint \frac{dz}{2 \pi i z} \frac{\prod_{i=1}^8 \Gamma(\sqrt[4]{p q}s_i z^{\pm};p,q)}{\Gamma(z^{\pm 2};p,q)},
\end{align}
where the path of the contour is taken to be the unit circle with positive orientation. The chemical potentials of $SU(8)$ group $s_i$ obey the balancing condition $\prod_{i=1}^8 s_i=1$.  In this theory we have a chiral scalar multiplet in the fundamental representations of $SU(2)$ and $SU(8)$. 

In the paper \cite{Spiridonov:2008zr}, Spiridonov and the second author established that there exist $71$ dual magnetic theories in addition to the above electric theory. They classified these $71$ theories in three groups. 

The first type of dual magnetic theory is the theory which was found by Csaki et al. in \cite{Csaki:1997cu}. There are 35 dual theories of this type and all of them are $4d$ $SU(2)$ ${\cal N}=1$ theories with $SU(4)_l \times SU(4)_r \times U(1)_B$ flavor group, two scalar multiplets in the fundamental representation, a gauge field in the adjoint representation of the gauge group, and two singlets in the antisymmetric tensor representations of $SU(4)$ group. The index for this type of theory is
\begin{align}
    I_M^{(1)} &=  \frac{(p;p)_{\infty} (q;q)_{\infty}}{2}
\prod_{1 \leq i < j \leq 4} \Gamma((pq)^{1/2} s_i s_j;p,q)
 \prod_{5 \leq i < j \leq 8} \Gamma((p q)^{1/2} s_i s_j;p,q) \nonumber \\  
& \quad \times \oint \frac{\prod_{i=1}^4  \Gamma((p q)^{1/4}v^{-2} s_i z^{\pm 1};p,q)
\prod_{i=5}^8  \Gamma((pq)^{1/4}v^2 s_i z^{\pm 1};p,q)}
{\Gamma(z^{\pm 2};p,q)}  \frac{d z}{2 \pi i z},
\end{align}
where $v$ is a chemical potential of $U(1)_B$
\begin{equation}
v=\sqrt[4]{s_1 s_2 s_3 s_4}, \qquad v^{-1}=\sqrt[4]{s_5 s_6 s_7 s_8} \; .
\end{equation}

The second type is the original Seiberg dual theory \cite{Seiberg:1994pq} with $SU(2)$ gauge group and $SU(4) \times SU(4) \times U(1)_B \times U(1)_R$ flavor group, one singlet in the fundamental representation of $SU(4)$ and all other matter content is the same as the theory above. The superconformal index for this theory is
\begin{align}
I_M^{(2)} &=\frac{ (p;p)_{\infty} (q;q)_{\infty}}{2}
\prod_{i=1}^4\prod_{j=5}^8 \Gamma((p q)^{1/2} s_i s_j;p,q) \nonumber \\ 
& \quad \times \oint \frac{\prod_{i=1}^4
\Gamma((pq)^{1/4} v^2 s_i^{-1} z^{\pm 1};p,q)
\prod_{i=5}^8  \Gamma((pq)^{1/4}
v^{-2} s_i^{-1} z^{\pm 1};p,q)}
{\Gamma(z^{\pm 2};p,q)} \frac{d z}{2 \pi i z} \;.
\end{align}

The theory considered by Intriligator and Pouliot in \cite{Intriligator:1995ne} corresponds to the third type. There is only a single model of this type and it has $SU(8)$ flavor group and $SU(2)$ gauge group, one chiral scalar multiplet in the fundamental representation of the gauge group and antisymmetric representation of the flavor group, a gauge field in the adjoint representation of the gauge group and one singlet in the antisymmetric tensor representation of flavor group. The superconformal index is
\begin{align}
I_M^{(3)} &=\frac{(p;p)_{\infty} (q;q)_{\infty}}{2}
\prod_{1 \leq i < j \leq 8} \Gamma((p q)^{1/2} s_i s_j;p,q)  \oint \frac{\prod_{i=1}^8  \Gamma((p q)^{1/4} s_i^{-1}
z^{\pm 1};p,q)}{\Gamma(z^{\pm 2};p,q)} \frac{d z}{2 \pi i  z}.
\end{align}

More detailed explanations about these dual theories can be found in the original paper \cite{Spiridonov:2008zr} and also in \cite{Khmelnitsky:2009vc}. The equality of all four indices follows from the following identity \cite{S2} 
\begin{align}\label{Sp}
& I(t_1, \ldots , t_8;p,q) = \prod_{1 \leq j < k \leq 4}
\Gamma(t_jt_k;p,q)\Gamma(t_{j+4}t_{k+4};p,q)\, I(s_1, \ldots , s_8;p,q),
\end{align}
where the complex variables $s_j,\, |s_j|<1,$ are given in terms of $t_j$ $(j=1, \ldots, 8)$,
\begin{align}
s_j &= \rho^{-1} t_j, \ j=1,2,3,4, \quad s_j = \rho t_j, \
j=5,6,7,8, \\ \nonumber 
\rho &= \sqrt{\frac{t_1t_2t_3t_4}{p q}}=\sqrt{\frac{p q}{t_5t_6t_7t_8}} \; .
\end{align}

All $72$ dual theories are associated with the orbit of  the $W(E_7)$ Weyl group. Using this fact Spiridonov and the second author speculated in \cite{Spiridonov:2008zr}, that the index may have global symmetry group $E_7$. In fact, Dimofte and Gaiotto explicitly showed in \cite{Dimofte:2012pd} that the theories in question, when coupled to $5d$ hypermultiplet, have an enhanced symmetry group $E_7$. In order to show this, they added the $5d$ hypermultiplet contributions with a specific boundary condition to the index
\begin{align} \label{4d5dindex}
I_{4d/5d,\; N_F=4} =\prod_{1\leq i<j \leq 8} \frac{1}{\left(\sqrt{p q} (s_i s_j)^{-1};p,q\right)_{\infty}} \frac{(p;p)_\infty (q;q)_\infty}{2}
 \oint \frac{dz}{2 \pi i z} \frac{\prod_{i=1}^8 \Gamma(\sqrt[4]{p q}s_i z^{\pm};p,q)}{\Gamma(z^{\pm 2};p,q)}.
\end{align}
where the term 
\begin{equation}
\prod_{1\leq i<j \leq 8} \frac{1}{\left(\sqrt{p q} (s_i s_j)^{-1};p,q\right)_{\infty}}
\end{equation}
corresponds to a $5d$ hypermultiplet \cite{Kim:2012gu}. By setting all chemical potentials to 1 and redefining $p=t^3 y$, $q=t^3 y^{-1}$ one can easily read off the $E_7$ symmetry of the index by expanding the last expression in powers of $t$ and $y$
\begin{equation}
I_{4d/5d,\; N_F=4}=1+56 t^3+1463 t^6+3002 t^9 y+\ldots \, ,
\end{equation}
where the coefficients $56$ and $1463$ are the dimensions of the irreducible representations of $E_7$ with highest weight $[0,0,0,0,0,0,1]$ and $[0,0,0,0,0,0,2]$, respectively and $3002=1539_{[0,0,0,0,0,1,0]}+1463_{[0,0,0,0,0,0,2]}$\footnote{To find dimensions of irreducible representations of Lie algebras one can use \\ \url{http://www-math.univ-poitiers.fr/~maavl/LiE/form.html}}. 

Remarkably, the new index is invariant under the transformation of the chemical potentials to their duals and the expression (\ref{Sp}) becomes \cite{Dimofte:2012pd}
\begin{align}
& I(t_1, \ldots , t_8;p,q) = I(s_1, \ldots , s_8;p,q).
\end{align}

If we set $s_7 s_8=\sqrt{p q}$ in (\ref{inindex}) one gets the reduction\footnote{We have used the reflection identity for an elliptic Gamma function $\Gamma(z;p,q) \Gamma(pq z^{-1};p,q)=1$.} of the index from $N_F=4$ to $N_F=3$.  When we apply this reduction for the integrals $I^{(1)}_M$ and $I^{(2)}_M$, setting $s_4 s_5=\sqrt{pq}$ and $s_7 s_8=\sqrt{pq}$, respectively, we end up with the flavor group $SU(3)_l \times SU(3)_r \times U(1)_B \times U(1)_{add}$ for $I^{(1)}_M$ and the flavor group $SU(4) \times SU(2) \times SU(2)_{add} \times U(1)_B$ for $I^{(2)}_M$. The observation that one gets additional symmetries such as $SU(2)_{add}$ and $U(1)_{add}$ in the reduced theories, suggests that the reduced theories may also have larger symmetry than $SU(6)$, in fact $E_6$ flavor symmetry. Indeed it is possible to show this by adding the 5d hypermultiplet contribution to the index and apply reduction procedure. The new reduced index is
\begin{align} \label{4d/5dNF=3}
I_{4d/5d,\; N_F=3} &=\prod_{1\leq i<j \leq 6} \frac{1}{\left((p q)^{\frac23}s_i^{-1} s_j^{-1};p,q\right)_{\infty}} \prod_{i=1}^6 \frac{1}{\left((p q)^{\frac13} s_i^{-1} w^{\pm 1};p,q \right)_{\infty}} \nonumber \\  
& \quad \times \frac{(p,p)_\infty (q,q)_\infty}{2} \oint \frac{dz}{2 \pi i z} \frac{\prod_{i=1}^6 \Gamma(\sqrt[6]{p q}s_i z^{\pm};p,q)}{\Gamma(z^{\pm 2};p,q)} \; .
\end{align}
Note that we have redefined the chemical potentials $s_i \rightarrow (p q)^{-1/12} s_i$. The balancing condition is $\prod_{i=1}^6 s_i=1$. Now by setting all chemical potentials to 1 and redefining $p=t^3 y$ and $q=t^3 y^{-1}$ one can read off the $E_6$ symmetry of the index
\begin{equation}
I_{4d/5d,\; N_F=3}=1+27 t^2+378 t^4+3653 t^6+27 t^5 (y^{-1}+y) + \ldots
\end{equation}
The coefficient $27$ is the dimension of the irreducible representation of $E_6$ with highest weight $[1,0,0,0,0,0]$ and
\begin{align}
& 378=351_{[0,0,1,0,0,0]}+27_{[1,0,0,0,0,0]}, \\
& 3653=3003_{[3,0,0,0,0,0]}+650_{[1,0,0,0,0,1]}.
\end{align}

\vspace{0.7cm}

There is a reduction scheme \cite{Dolan:2011rp} (also see \cite{Gadde:2011ia,Imamura:2011uw}) of the superconformal index for a $4d$ supersymmetric theory  to the partition function for a $3d$ theory. Actually from the mathematical point of view this reduction nothing but a special limit that brings elliptic gamma functions to the hyperbolic level. Let us do this procedure for the index (\ref{4d/5dNF=3}), following \cite{Dolan:2011rp}. First we reparameterize
\begin{equation}
p=e^{2\pi i v \omega_1}, \;\; q=e^{2 \pi i v \omega_2}, \;\; z=e^{2 \pi i v u}, \;\; s_i=e^{2\pi i v \alpha_i}, \;\; w=e^{2 \pi i v \alpha_7} \; ,
\end{equation}
and use the asymptotic formula for the elliptic $\Gamma$-functions. Recall that in the limit $v\rightarrow 0$ the elliptic $\Gamma$-function reduce to hyperbolic $\gamma^{(2)}(z)$-function
\begin{equation}
\Gamma(e^{2 \pi \textup{i} v z};e^{2 \pi \textup{i} v \omega_1}, e^{2 \pi
\textup{i} v \omega_2}) \stackrel[{v \rightarrow 0}]{}{=} e^{-\pi
\textup{i}(2z-(\omega_1+\omega_2))/24 v\omega_1\omega_2} \gamma^{(2)}(z;\omega_1,\omega_2) \; ,
\end{equation}
where
\begin{equation}
\gamma^{(2)}(u;\omega_1, \omega_2)=e^{-\pi i B_{2,2}(u;\mathbf{\omega})/2} \frac{(e^{2 \pi i u/\omega_1}\tilde{q};\tilde{q})}{(e^{2 \pi i u/\omega_1};q)} \quad \text{with} \quad q=e^{2 \pi i \omega_1/\omega_2}, \quad \tilde{q}=e^{-2 \pi i \omega_2/\omega_1} \; ,
\end{equation}
and  $B_{2,2}(u;\mathbf{\omega})$ is the second order Bernoulli polynomial,
\begin{equation} B_{2,2}(u;\mathbf{\omega}) =
\frac{u^2}{\omega_1\omega_2} - \frac{u}{\omega_1} -
\frac{u}{\omega_2} + \frac{\omega_1}{6\omega_2} +
\frac{\omega_2}{6\omega_1} + \frac 12.
\end{equation}
In the limit $v \rightarrow 0$ we also have
\begin{equation}
(z;p,q)_\infty \stackrel[{v \rightarrow 0}]{}{\rightarrow} \frac{1}{\Gamma_2(u; \omega_1,\omega_2)} \; ,
\end{equation}
where $\Gamma_2(u; \omega_1,\omega_2)$ is the Barnes double Gamma function (see Appendix A).

To go further let us apply the limit $v \rightarrow 0$ to the index (\ref{4d/5dNF=3}) and use the asymptotic formulae above. Finally we arrive at\footnote{We have also used the reflection identity and some asymptotic formulas for $\gamma^{(2)}(z)$ function (see Appendix B). Here and below we will use the shorthand notations $
\gamma^{(2)}(a,b;\omega_1,\omega_2) \equiv
\gamma^{(2)}(a;\omega_1,\omega_2) \gamma^{(2)}(b;\omega_1,\omega_2),
$
and
$
\gamma^{(2)}(a\pm u;\omega_1,\omega_2) \equiv
\gamma^{(2)}(a+u;\omega_1,\omega_2)
\gamma^{(2)}(a-u;\omega_1,\omega_2)
$.}
%Using all these asymptotic expressions one obtains 
\begin{equation}
I_{4d/5d} \stackrel[{v \rightarrow 0}]{}{=}  e^{\pi i (\omega_1+\omega_2)/12 v\omega_1\omega_2} I_{4d/5d}^r \; ,
\end{equation}
where
\begin{align}
I_{4d/5d}^r &=\prod_{1\leq i<j \leq 6} \Gamma_2\big(\frac{\omega_1+\omega_2}{2}-(\alpha_i+\alpha_j)\big) \prod_{i=1}^6 \Gamma_2\big(-\frac{\omega_1+\omega_2}{2}-(\alpha_i \pm \alpha_7)\big)  \nonumber \\
 & \quad \times \frac12 \int \frac{d u}{i \sqrt{\omega_1 \omega_2}} \frac{\prod_{i=1}^6 \gamma^{(2)}(\alpha_i \pm u+\frac{\omega_1+\omega_2}{4};\omega_1,\omega_2)}{\gamma^{(2)}(\pm 2 u;\omega_1,\omega_2)}.
\end{align}
If one considers 
\begin{equation}
\alpha_5=\xi_1+a S, \quad \quad \alpha_6=\xi_2-a S,
\end{equation}
and applies the additional limit $S \rightarrow \infty$, then the final answer gives an expression for the partition function of $3d$ ${\cal N}=2$ SYM theory\footnote{For $\omega_1=\frac{1}{\omega_2}$ we obtained that \begin{align} \nonumber
F&=\left(-\xi_1-\frac{5 i \pi  \xi_1 }{6}-\xi_ 2-\frac{i \pi  \xi_ 2}{6}\right) (\omega+\frac{1}{\omega})+(\frac{i \pi }{3}-\frac43)(\frac{1}{\omega}+\omega)^2- i \pi\\ \nonumber
& \quad -\frac{5}{2} i \pi  \xi_1^2+\frac{15 \xi_2^2}{2}+\left(\frac{3}{2}-\frac{i \pi }{2}\right)\left( \alpha_1^2+\alpha_2^2+\alpha_3^2+\alpha_4^2+5 \xi_1^2-2\xi_1 \xi_2+8 \alpha_7^2 \right).
\end{align}} \cite{Jafferis:2010un, Hama:2010av, Hama:2011ea}
\begin{equation}
Z_{3d/4d} \stackrel[{S \rightarrow \infty}]{}{\approx}  F Z_{3d/4d}^r \;,
\end{equation}
where
\begin{align} \label{partfunc}
Z_{3d/4d}^r &= \Gamma_2(\frac{\omega_1+\omega_2}{2}-\xi_1-\xi_2) \prod_{1\leq i<j \leq 4} \Gamma_2\big(\frac{\omega_1+\omega_2}{2}-(\alpha_i+\alpha_j)\big) \prod_{i=1}^4 \Gamma_2\big(-\frac{\omega_1+\omega_2}{2}-(\alpha_i \pm \alpha_7)\big)  \nonumber \\
 & \quad \times \frac12 \int \frac{d u}{i \sqrt{\omega_1 \omega_2}} \frac{\prod_{i=1}^4 \gamma^{(2)}(\alpha_i \pm u+\frac{\omega_1+\omega_2}{4};\omega_1,\omega_2)}{\gamma^{(2)}(\pm 2 u;\omega_1,\omega_2)}.
\end{align}
From the physical point of view this reduction corresponds to adding mass terms to two quark supermultiplets and then integrating them out by sending their masses to infinity. As one can see, this theory has $4$ quarks, one chiral field in the antisymmetric representation of the gauge group, and contributions from a $5d$ hypermultiplet.

\section{Reduction to $N_f=4$}

In \cite{Dimofte:2012pd} it was shown that $3d$ ${\cal N}=2$ theory with $N_f=6$ has $SO(12)$ symmetry. The authors obtained the index of the $3d$ theory by reduction from $4d$ ${\cal N}=1$ theory with $N_F=4$ inspired by \cite{Teschner:2012em}.  We will now demonstrate that the index for the 3d ${\cal N}=2$ SYM theory with $4$ quarks has $SO(10)$ symmetry group. 

The expression for the index of the electric 3d ${\cal N}=2$ supersymmetric theory \cite{Kim:2009wb,Imamura:2011su} with an arbitrary number of flavors $N_f$ and chemical potentials $s_i$, $t_i$, $(i=1,...,N_f)$ is \cite{Krattenthaler:2011da}
\begin{align}
I_{3d, N_f} &=\prod_{a,b=1}^{N_f}\frac{1}{\big(q^{\frac12} t_a^{-1} s_b^{-1};q\big)_\infty} \sum_{k \in \mathbb{Z}} a^{N_f|k|/2} \nonumber \\
& \quad \times \oint \frac{dz}{2 \pi \textup{i} z} \prod_{i=1}^{N_f} \frac{(a^{1/2}q^{1/2+|k|/2} t_i^{-1}
z;q)_\infty}{(a^{-1/2}q^{1/2+|k|/2} t_i z^{-1};q)_\infty}  \frac{(a^{1/2}q^{1/2+|k|/2} s_i^{-1}
z^{-1};q)_\infty}{(a^{-1/2}q^{1/2+|k|/2} s_i z;q)_\infty},
\end{align}
where $\prod_{a=1}^{N_f} t_a=1$ and $\prod_{a=1}^{N_f} s_a=1$. It is clear that by taking $a=q^{\frac12}$ for $N_f=4$ ($8$ quarks), we obtain the following expression 
\begin{align}
I_{3d, N_f=6} &=\prod_{a,b=1}^{4}\frac{1}{\big(q^{\frac12} t_a^{-1} s_b^{-1};q\big)_\infty} \sum_{k \in \mathbb{Z}}  q^{|k|} \nonumber \\
& \quad \times \oint \frac{dz}{2 \pi \textup{i} z}
\prod_{i=1}^{4} \frac{( q^{1/4}q^{1/2+|k|/2} t_i^{-1}
z;q)_\infty}{(q^{-1/4}q^{1/2+|k|/2} t_i z^{-1};q)_\infty}  \frac{(q^{1/4}q^{1/2+|k|/2} s_i^{-1}
z^{-1};q)_\infty}{(q^{-1/4}q^{1/2+|k|/2} s_i z;q)_\infty} \; .
\end{align}
One can rewrite this index in the following form \cite{Dimofte:2012pd} 
\begin{align} \label{2repres}
I_{3d, N_f=6}&=\frac{1}{\big(q^{\frac12}f_1 f_2 f_3 f_4 f_5 f_6;q\big)_\infty} \prod_{1\leq i<j\leq 6} \frac{1}{\big(q^{\frac12} f_i^{-1} f_j^{-1}; q \big)_\infty} \nonumber \\
  &\quad \times \frac12 \sum_{k\in \mathbb{Z}}\oint \frac{d z}{2\pi i z} (1-q^{|k|}z^{\pm 2}) \prod_{i=1}^6 f_i^{-|k|}   \frac{1-q^{r+\frac12 |k|+1} (q^\frac14 f_i z^{\pm 1})^{-1}}{1-q^{r+\frac12 |k|}q^\frac14 f_i z^{\pm 1}}, 
\end{align}
where $f_i=t_i/\sqrt{t_1 t_2 t_3 s_4}$ and $f_{i+3}=s_i \sqrt{t_1 t_2 t_3 s_4}$ $(i=1,2,3)$. The reduction of superconformal indices in $3d$ is similar to the $4d$ case. For the result of this paper, we set $f_5 f_6=q^{\frac12}$ which reduces the index of the theory with $6$ quarks to the index of the theory with $4$ quarks
\begin{align}
I_{3d, N_f=4} &=\frac{(q^{1/3};q)_{\infty}}{\big(q f_1 f_2 f_3 f_4;q\big)_\infty}\prod_{1\leq i<j\leq 4} \frac{1}{\big(q^{\frac12} f_i^{-1} f_j^{-1};q \big)_\infty} \prod_{i=1}^4 \frac{1}{\big(q^{\frac12} f_i^{-1} q^{-\frac14}v^{ \pm 1};q\big)_\infty} \nonumber \\
&\quad \times \frac12 \sum_{k\in \mathbb{Z}}\oint \frac{d z}{2\pi i z} (1-q^{|k|}z^{\pm 2}) \prod_{i=1}^4 f_i^{-|k|}    \frac{1-q^{r+\frac12 |k|+\frac34} f_i^{-1} z^{\pm 1}}{1-q^{r+\frac12 |k|+\frac14} f_i z^{\pm 1}} \; ,
\end{align}
where the term $(q^{\frac13};q)_{\infty}$ is a monopole contribution. Note that we have chosen the representation (\ref{2repres}) of the index because it is closely related to the $3d$ ${\cal N}=2$ partition function (\ref{partfunc}). This procedure can be repeated for the initial expression of the index (\ref{2repres}) in a similar way. Now one can read off the $SO(10)$--invariant operator content of the theory by expanding the last expression in powers of $q$ and setting all chemical potentials to $1$
\begin{equation}
I=1+16 q^{1/3}+136 q^{2/3}+816 q+3892 q^{4/3}+\ldots
\end{equation}
The coefficients are related to the dimensions of irreducible representations of $SO(10)$
\begin{align}
& 16 ~~ \text{is the dimension of the spinor representation of $SO(10)$}\\
& 136=54_{[2,0,0,0,0]}+45_{[0,1,0,0,0]}+16_{[0,0,0,1,0]}+10_{[1,0,0,0,0]}+1_{[0,0,0,0,0]}, \\
& 816=320_{[1,1,0,0,0]}+210_{[0,0,0,1,1]}+144_{[1,0,0,1,0]}+126_{[0,0,0,2,0]}+16_{[0,0,0,1,0]},\\
& 3892=2772_{[0,0,0,4,0]}+945_{[1,0,1,0,0]}+120_{[0,0,1,0,0]}+54_{[2,0,0,0,0]}+1_{[0,0,0,0,0]}.
%& 15680=9504_{[0,0,0,5,0]}+5280_{[1,0,03,0]}+770_{[0,2,0,0,0]}+126_{[0,0,0,2,0]}
\end{align}

\vspace{0.4cm}

\noindent \textbf{Acknowledgments.} IG would like to thank Jan Plefka for his support and is also grateful to Harald Dorn for pleasant discussions. The authors especially thank Ben Hoare for proof-reading the manuscript and suggesting valuable improvements for it.

\appendix
\section*{Appendices}

\section{Barnes double Gamma function}

The Barnes double Gamma function $\Gamma_2(u; \omega_1,\omega_2)$ is defined as
\begin{equation}
\log \Gamma_2(x;a,b)=\zeta'_{2}(0;a,b,x)+\log \rho_2 (a,b),
\end{equation}
where 
\begin{align}
\zeta_{2}(s;a,b,x)&=\sum_{m,n=0} (a m+b n+x)^{-s}\\
\log \rho_2 (a,b) &=- \lim_{x\rightarrow 0} \left(\zeta'_2 (0;a,b,x)+\log x\right)
\end{align}
There is also the integral representation of this function 
\begin{equation}
\Gamma_2(x;a,b)=\exp \left(\frac{1}{2 \pi i}\int_{C_H} \frac{e^{-x t} (\log (-t)+\gamma)}{t(1-e^{-at})(1-e^{-bt})}dt \right),
\end{equation}
where $\gamma$ is the Euler constant and the Hankel contour $C_H$ starts and finishes near the point $+\infty$, turning around the half--axis $[0,\infty)$ anticlockwise.

Useful references for specific details are \cite{Spiridonov,Spreafico}.

\section{Hyperbolic gamma-function}

The reflection identity for a hyperbolic gamma-function is as follows
\begin{equation}
\gamma^{(2)}(z,\omega_1+\omega_2-z;\omega_1,\omega_2) = 1,
\end{equation} 
and the asymptotic formulas are
\begin{align}
\lim_{u \rightarrow \infty}
e^{\frac{\pi \textup{i}}{2} B_{2,2}(u;\omega_1,\omega_2)} \gamma^{(2)}(u; \omega_1,\omega_2)
& =  1, \text{ \ \ for } \text{arg }\omega_1 < \text{arg } u < \text{arg }\omega_2 + \pi,  \\
\lim_{u \rightarrow \infty}e^{-\frac{\pi \textup{i}}{2} B_{2,2}(u;\omega_1,\omega_2)} \gamma^{(2)}(u;\omega_1,\omega_2)
& =  1, \text{  \ \ for } \text{arg } \omega_1 - \pi < \text{arg } u < \text{arg }\omega_2.
\end{align}


\begin{thebibliography}{99} 

%\cite{Romelsberger:2007ec}
\bibitem{Romelsberger:2007ec}
  C.~Romelsberger,
  ``Calculating the Superconformal Index and Seiberg Duality,''
  \href{http://arxiv.org/abs/0707.3702}{arXiv:0707.3702}.
  %%CITATION = ARXIV:0707.3702;%%

%\cite{Dolan:2008qi}
\bibitem{Dolan:2008qi}
  F.~A.~Dolan and H.~Osborn,
  ``Applications of the Superconformal Index for Protected Operators and q-Hypergeometric Identities to N=1 Dual Theories,''
  Nucl.\ Phys.\ B {\bf 818} (2009) 137
  [\href{http://arxiv.org/abs/0801.4947}{arXiv:0801.4947}].
  %%CITATION = ARXIV:0801.4947;%%

%\cite{Spiridonov:2008zr}
\bibitem{Spiridonov:2008zr}
  V.~P.~Spiridonov and G.~S.~Vartanov,
  ``Superconformal indices for N = 1 theories with multiple duals,''
  Nucl.\ Phys.\ B {\bf 824} (2010) 192
  [\href{http://arxiv.org/abs/0811.1909}{arXiv:0811.1909}].
  %%CITATION = ARXIV:0811.1909;%%
  
%\cite{Spiridonov:2009za}
\bibitem{Spiridonov:2009za}
  V.~P.~Spiridonov and G.~S.~Vartanov,
  ``Elliptic Hypergeometry of Supersymmetric Dualities,''
  Commun.\ Math.\ Phys.\  {\bf 304} (2011) 797
  [\href{http://arxiv.org/abs/0910.5944}{arXiv:0910.5944}].
  %%CITATION = ARXIV:0910.5944;%%

%\cite{Spiridonov:2011hf}
\bibitem{Spiridonov:2011hf}
  V.~P.~Spiridonov and G.~S.~Vartanov,
  ``Elliptic hypergeometry of supersymmetric dualities II. Orthogonal groups, knots, and vortices,''
  \href{http://arxiv.org/abs/0811.1909}{arXiv:1107.5788}.
  %%CITATION = ARXIV:1107.5788;%%
  %26 citations counted in INSPIRE as of 05 Mar 2013

%\cite{Gadde:2009kb}
\bibitem{Gadde:2009kb}
  A.~Gadde, E.~Pomoni, L.~Rastelli and S.~S.~Razamat,
  ``S-duality and 2d Topological QFT,''
  JHEP {\bf 1003} (2010) 032
  [\href{http://arxiv.org/abs/0910.2225}{arXiv:0910.2225}].
  %%CITATION = ARXIV:0910.2225;%%
  
%\cite{Gadde:2011uv}
\bibitem{Gadde:2011uv}
  A.~Gadde, L.~Rastelli, S.~S.~Razamat and W.~Yan,
  ``Gauge Theories and Macdonald Polynomials,''
  \href{http://arxiv.org/abs/1110.3740}{arXiv:1110.3740}.
  %%CITATION = ARXIV:1110.3740;%%
  
%\cite{Spiridonov:2010qv}
\bibitem{Spiridonov:2010qv}
  V.~P.~Spiridonov and G.~S.~Vartanov,
  ``Superconformal indices of ${\mathcal N}=4$ SYM field theories,''
  Lett.\ Math.\ Phys.\  {\bf 100} (2012) 97
  [\href{http://arxiv.org/abs/1005.4196}{arXiv:1005.4196}].
  %%CITATION = ARXIV:1005.4196;%%

\bibitem{S1}  
V. P. Spiridonov,
``On the elliptic beta function'', Uspekhi Mat. Nauk {\bf 56}
(1) (2001), 181--182 (Russian Math. Surveys {\bf 56} (1) (2001),
185--186).

%\cite{S2}
\bibitem{S2} 
V.~P.~Spiridonov, 
``Theta hypergeometric integrals,'' 
Algebra i Analiz {\bf 15} (2003) no. 6, 161--215 (St.
Petersburg Math. J. {\bf 15} (2003) 929--967), [\href{http://arxiv.org/abs/math/0303205}{math.CA/0303205}].

%\cite{Romelsberger:2005eg}
\bibitem{Romelsberger:2005eg}
  C.~Romelsberger,
  ``Counting chiral primaries in N = 1, d=4 superconformal field theories,''
  Nucl.\ Phys.\ B {\bf 747} (2006) 329
  [\href{http://arxiv.org/abs/hep-th/0510060}{hep-th/0510060}].
  %%CITATION = HEP-TH/0510060;%%

%\cite{Kinney:2005ej}
\bibitem{Kinney:2005ej}
  J.~Kinney, J.~M.~Maldacena, S.~Minwalla and S.~Raju,
  ``An Index for 4 dimensional super conformal theories,''
  Commun.\ Math.\ Phys.\  {\bf 275} (2007) 209
  [\href{http://arxiv.org/abs/hep-th/0510251}{hep-th/0510251}].
  %%CITATION = HEP-TH/0510251;%%

%\cite{Witten:1982df}
\bibitem{Witten:1982df}
  E.~Witten,
  ``Constraints on Supersymmetry Breaking,''
  Nucl.\ Phys.\ B {\bf 202} (1982) 253.
  %%CITATION = NUPHA,B202,253;%%

%\cite{Festuccia:2011ws}
\bibitem{Festuccia:2011ws}
  G.~Festuccia and N.~Seiberg,
  ``Rigid Supersymmetric Theories in Curved Superspace,''
  JHEP {\bf 1106} (2011) 114
  [\href{http://arxiv.org/abs/1005.4196}{arXiv:1105.0689}].
  %%CITATION = ARXIV:1105.0689;%%
  %85 citations counted in INSPIRE as of 05 Mar 2013


%\cite{Sudano:2011aa}
\bibitem{Sudano:2011aa}
  M.~Sudano,
  ``The Romelsberger Index, Berkooz Deconfinement, and Infinite Families of Seiberg Duals,''
  JHEP {\bf 1205} (2012) 051
  [\href{http://arxiv.org/abs/1112.2996}{arXiv:1112.2996}].
  %%CITATION = ARXIV:1112.2996;%%
  
%\cite{Feng:2007ur}
\bibitem{Feng:2007ur}
  B.~Feng, A.~Hanany and Y.~-H.~He,
  ``Counting gauge invariants: The Plethystic program,''
  JHEP {\bf 0703} (2007) 090
  [\href{http://arxiv.org/abs/hep-th/0701063}{hep-th/0701063}].
  %%CITATION = HEP-TH/0701063;%%

%\cite{Spiridonov}
\bibitem{Spiridonov}
V.~P.~Spiridonov, 
``Essays on the theory of elliptic hypergeometric functions'',
Uspekhi Mat. Nauk, 63:3(381) (2008), 372 

%\cite{Dimofte:2012pd}
\bibitem{Dimofte:2012pd}
  T.~Dimofte and D.~Gaiotto,
  ``An E7 Surprise,''
  JHEP {\bf 1210} (2012) 129
  [\href{http://arxiv.org/abs/1209.1404}{arXiv:1209.1404}].
  %%CITATION = ARXIV:1209.1404;%%
  %4 citations counted in INSPIRE as of 13 Feb 2013


%\cite{Csaki:1997cu}
\bibitem{Csaki:1997cu}
  C.~Csaki, M.~Schmaltz, W.~Skiba and J.~Terning,
  ``Selfdual N=1 SUSY gauge theories,''
  Phys.\ Rev.\ D {\bf 56} (1997) 1228
  [\href{http://arxiv.org/abs/hep-th/9701191}{hep-th/9701191}].
  %%CITATION = HEP-TH/9701191;%%
  
%\cite{Seiberg:1994pq}
\bibitem{Seiberg:1994pq}
  N.~Seiberg,
  ``Electric - magnetic duality in supersymmetric nonAbelian gauge theories,''
  Nucl.\ Phys.\ B {\bf 435} (1995) 129
  [\href{http://arxiv.org/abs/hep-th/9411149}{hep-th/9411149}].
  %%CITATION = HEP-TH/9411149;%%

%\cite{Intriligator:1995ne}
\bibitem{Intriligator:1995ne}
  K.~A.~Intriligator and P.~Pouliot,
  ``Exact superpotentials, quantum vacua and duality in supersymmetric SP(N(c)) gauge theories,''
  Phys.\ Lett.\ B {\bf 353} (1995) 471
  [\href{http://arxiv.org/abs/hep-th/9505006}{hep-th/9505006}].
  %%CITATION = HEP-TH/9505006;%%
  
%\cite{Khmelnitsky:2009vc}
\bibitem{Khmelnitsky:2009vc}
  A.~Khmelnitsky,
  ``Interpreting multiple dualities conjectured from superconformal index identities,''
  JHEP {\bf 1003} (2010) 065
  [\href{http://arxiv.org/abs/0912.4523}{arXiv:0912.4523}].
  %%CITATION = ARXIV:0912.4523;%%
  
%\cite{Kim:2012gu}
\bibitem{Kim:2012gu}
  H.~-C.~Kim, S.~-S.~Kim and K.~Lee,
  ``5-dim Superconformal Index with Enhanced En Global Symmetry,''
  JHEP {\bf 1210} (2012) 142
  [\href{http://arxiv.org/abs/0912.4523}{arXiv:1206.6781}].
  %%CITATION = ARXIV:1206.6781;%%
  %23 citations counted in INSPIRE as of 03 Mar 2013

%\cite{Dolan:2011rp}
\bibitem{Dolan:2011rp}
  F.~A.~H.~Dolan, V.~P.~Spiridonov and G.~S.~Vartanov,
  ``From 4d superconformal indices to 3d partition functions,''
  Phys.\ Lett.\ B {\bf 704} (2011) 234
  [\href{http://arxiv.org/abs/1104.1787}{arXiv:1104.1787}].
  %%CITATION = ARXIV:1104.1787;%%

%\cite{Gadde:2011ia}
\bibitem{Gadde:2011ia}
  A.~Gadde and W.~Yan,
  ``Reducing the 4d Index to the $S^3$ Partition Function,''
  JHEP {\bf 1212} (2012) 003
  [\href{http://arxiv.org/abs/1104.2592}{arXiv:1104.2592}].
  %%CITATION = ARXIV:1104.2592;%%
  %32 citations counted in INSPIRE as of 14 Feb 2013

%\cite{Imamura:2011uw}
\bibitem{Imamura:2011uw}
  Y.~Imamura,
  ``Relation between the 4d superconformal index and the $S^3$ partition function,''
  JHEP {\bf 1109} (2011) 133
  [\href{http://arxiv.org/abs/1104.4482}{arXiv:1104.4482}].
  %%CITATION = ARXIV:1104.4482;%%
  %30 citations counted in INSPIRE as of 14 Feb 2013
  
%\cite{Jafferis:2010un}
\bibitem{Jafferis:2010un}
  D.~L.~Jafferis,
  ``The Exact Superconformal R-Symmetry Extremizes Z,''
  JHEP {\bf 1205} (2012) 159
  [\href{http://arxiv.org/abs/1104.4482}{arXiv:1012.3210}].
  %%CITATION = ARXIV:1012.3210;%%
  %143 citations counted in INSPIRE as of 03 Mar 2013

%\cite{Hama:2010av}
\bibitem{Hama:2010av}
  N.~Hama, K.~Hosomichi and S.~Lee,
  ``Notes on SUSY Gauge Theories on Three-Sphere,''
  JHEP {\bf 1103} (2011) 127
  [\href{http://arxiv.org/abs/1104.4482}{arXiv:1012.3512}].
  %%CITATION = ARXIV:1012.3512;%%
  %102 citations counted in INSPIRE as of 03 Mar 2013

%\cite{Hama:2011ea}
\bibitem{Hama:2011ea}
  N.~Hama, K.~Hosomichi and S.~Lee,
  ``SUSY Gauge Theories on Squashed Three-Spheres,''
  JHEP {\bf 1105} (2011) 014
  [\href{http://arxiv.org/abs/1104.4482}{arXiv:1102.4716}].
  %%CITATION = ARXIV:1102.4716;%%
  %73 citations counted in INSPIRE as of 03 Mar 2013


%\cite{Teschner:2012em}
\bibitem{Teschner:2012em}
  J.~Teschner and G.~S.~Vartanov,
  ``6j symbols for the modular double, quantum hyperbolic geometry, and supersymmetric gauge theories,''
  \href{http://arxiv.org/abs/1202.4698}{arXiv:1202.4698}.
  %%CITATION = ARXIV:1202.4698;%%
  
%\cite{Kim:2009wb}
\bibitem{Kim:2009wb}
  S.~Kim,
  ``The Complete superconformal index for N=6 Chern-Simons theory,''
  Nucl.\ Phys.\ B {\bf 821} (2009) 241
   [Erratum-ibid.\ B {\bf 864} (2012) 884]
  [\href{http://arxiv.org/abs/0903.4172}{arXiv:0903.4172}].
  %%CITATION = ARXIV:0903.4172;%%
  %95 citations counted in INSPIRE as of 20 Feb 2013
  
%\cite{Imamura:2011su}
\bibitem{Imamura:2011su}
  Y.~Imamura and S.~Yokoyama,
  ``Index for three dimensional superconformal field theories with general R-charge assignments,''
  JHEP {\bf 1104} (2011) 007
  [\href{http://arxiv.org/abs/0903.4172}{arXiv:1101.0557}].
  %%CITATION = ARXIV:1101.0557;%%
  %50 citations counted in INSPIRE as of 03 Mar 2013

%\cite{Krattenthaler:2011da}
\bibitem{Krattenthaler:2011da}
  C.~Krattenthaler, V.~P.~Spiridonov and G.~S.~Vartanov,
  ``Superconformal indices of three-dimensional theories related by mirror symmetry,''
  JHEP {\bf 1106} (2011) 008
  [\href{http://arxiv.org/abs/1103.4075}{arXiv:1103.4075}].
  %%CITATION = ARXIV:1103.4075;%%

%\cite{Spreafico}  
\bibitem{Spreafico} 
M.~Spreafico, 
``On the Barnes double zeta and Gamma functions,''
 Journal of Number Theory 129 (2009) 2035-2063.


  
\end{thebibliography}
\end{document}